\documentclass[runningheads]{llncs}
\usepackage{graphicx}
\usepackage[misc,geometry]{ifsym}
\usepackage{enumitem}
\usepackage{hyperref}
\usepackage{multirow}
\usepackage{array}
\usepackage{subfigure}
\usepackage{booktabs}
\usepackage{xcolor}
\usepackage{amsmath,bm,amssymb}
\usepackage{makecell}

\begin{document}
\title{OptMSM: Optimizing Multi-Scenario Modeling for Click-Through Rate Prediction\thanks{The first three authors contribute equally to this work.}}
\titlerunning{Optimizing Multi-Scenario Modeling for CTR Prediction}
\author{Xing Tang\inst{1} \and
Yang Qiao\inst{1} \and
Yuwen Fu\inst{1} \and
Fuyuan Lyu\inst{2} \and
Dugang Liu\inst{3}\textsuperscript{(\Letter)} \and
Xiuqiang He\inst{1}\textsuperscript{(\Letter)}
}
\authorrunning{Tang et al.}
\institute{
FiT, Tencent, Shenzhen, China\\
\email{\{shawntang,sunnyqiao,evenfu,xiuqianghe\}@tencent.com} \and
School of Computer Science, McGill University, Montreal, Canada \and
Guangdong Laboratory of Artificial Intelligence and Digital Economy (SZ), Shenzhen University, Shenzhen, China\\
% \email{dugang.ldg@gmail.com}
}

\toctitle{OptMSM: Optimizing Multi-Scenario Modeling for Click-Through Rate Prediction}
\tocauthor{Xing Tang, Yang Qiao, Yuwen Fu, Fuyuan Lyu, Dugang Liu, Xiuqiang He}
\maketitle              % typeset the header of the contribution

\begin{abstract}
% ==
A large-scale industrial recommendation platform typically consists of multiple associated scenarios, requiring a unified click-through rate (CTR) prediction model to serve them simultaneously.
% ==
Existing approaches for multi-scenario CTR prediction generally consist of two main modules: i) a scenario-aware learning module that learns a set of multi-functional representations with scenario-shared and scenario-specific information from input features, and ii) a scenario-specific prediction module that serves each scenario based on these representations.
% ==
However, most of these approaches primarily focus on improving the former module and neglect the latter module.
% ==
This can result in challenges such as increased model parameter size, training difficulty, and performance bottlenecks for each scenario.
% ==
To address these issues, we propose a novel framework called OptMSM (\textbf{Opt}imizing \textbf{M}ulti-\textbf{S}cenario \textbf{M}odeling).
% ==
First, we introduce a simplified yet effective scenario-enhanced learning module to alleviate the aforementioned challenges.
% ==
Specifically, we partition the input features into scenario-specific and scenario-shared features, which are mapped to specific information embedding encodings and a set of shared information embeddings, respectively.
% ==
By imposing an orthogonality constraint on the shared information embeddings to facilitate the disentanglement of shared information corresponding to each scenario, we combine them with the specific information embeddings to obtain multi-functional representations.
% ==
Second, we introduce a scenario-specific hypernetwork in the scenario-specific prediction module to capture interactions within each scenario more effectively, thereby alleviating the performance bottlenecks.
% ==
Finally, we conduct extensive offline experiments and an online A/B test to demonstrate the effectiveness of OptMSM. 
% ==
% Furthermore, we deploy OptMSM on a large-scale online financial recommendation platform, and online A/B tests show its significant performance advantages.
\end{abstract}
\section{Introduction}\label{sec:intro}
\begin{figure}[htbp]
    \centering
    \subfigure[Homepage]{
        \includegraphics[width=0.2\textwidth]{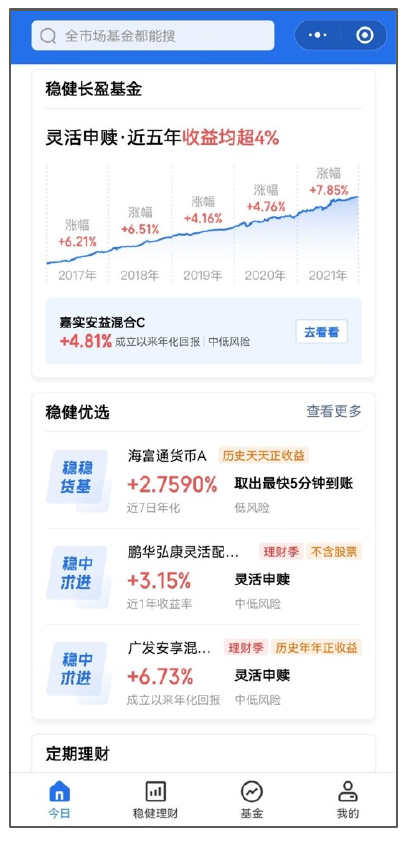}
    }
    \subfigure[BIP page]{
        \includegraphics[width=0.2\textwidth]{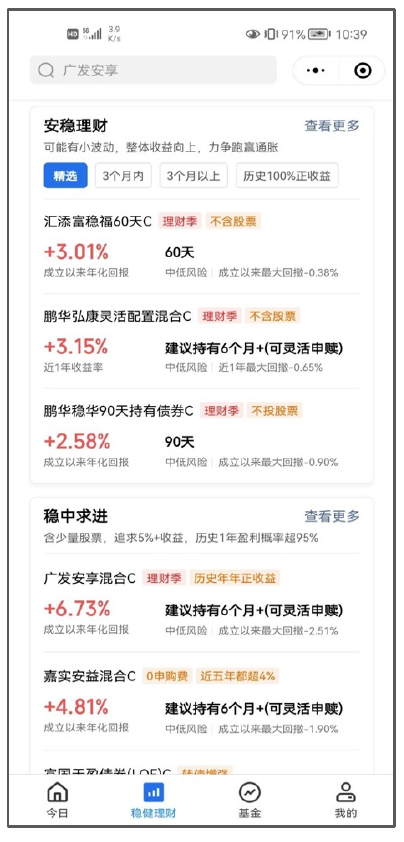}
    }
    \subfigure[AIP page]{
        \includegraphics[width=0.2\textwidth]{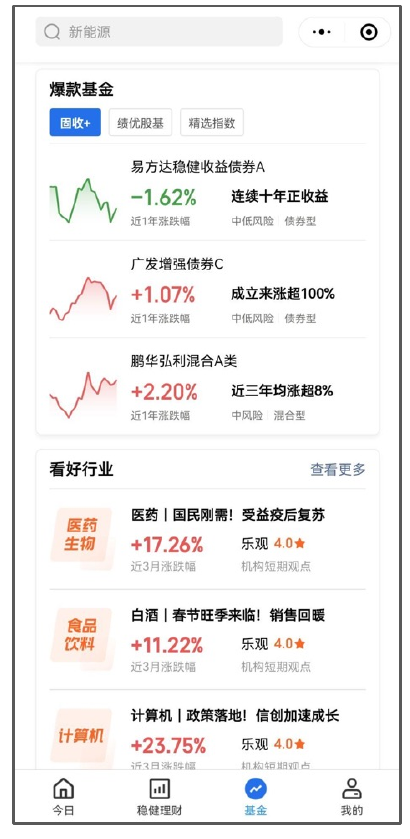}
    }
    \caption{The scenarios in Tencent Licaitong financial recommendation platform.}
    \label{fig:example}
\end{figure}

% ==
Click-through rate (CTR) prediction is a crucial component in online recommendation platform~\cite{Wide_Deep,DLRM,ADS,LR}, which aims to predict the probability of candidate items being clicked and return top-ranked items for each user. 
% ==
In practice, a business is usually divided into different scenarios based on different user groups or item categories~\cite{MRMD,MARL,sharing}, and the resource overhead of customizing a proprietary CTR prediction model for each scenario is too high.
% ==
Therefore, designing and deploying a unified CTR prediction model to efficiently serve all scenarios is a realistic challenge for a large-scale industrial recommendation platform.
% ==
Taking the Tencent Licaitong financial recommendation platform used in the online experiment as an example, as shown in Fig.~\ref{fig:example}, these scenarios include the homepage (HP), balanced investment portfolio page (BIP), and aggressive investment portfolio page (AIP).
% ==
% Specifically, HP is the first page that each user interacts with, where the users usually browse the items without a specific intent and the categories of items are mixed, and the BIP and AIP pages list the items with corresponding categories for the users with different specific intents, respectively.
Specifically, HP is the first page that each user interacts with, where the users usually browse the items without specific intent. The categories of items are mixed. The BIP and AIP pages list the items with corresponding categories for the users with different specific intents, respectively.
% ==
We focus on how to effectively utilize all the user interactions in multiple scenarios to obtain a desired CTR prediction model.

% ==
Different from single-scenario modeling~\cite{CTR_Survey}, multi-scenario modeling (MSM) for CTR prediction is proposed in previous works to address the above goals.
% ==
Existing works for MSM usually adopt the idea of multi-task learning to model the relationship between different scenarios~\cite{mtl,MMOE,ple}. They usually contain two main modules, i.e., the scenario-aware learning module and the scenario-specific prediction module.
% ==
The former is used to learn versatile scenario-aware representations, where scenario-shared and scenario-specific information are captured simultaneously. The latter uses a scenario-specific architecture to predict the corresponding scenario based on scenario-aware representations.
% ==
Obviously, the scenario-aware learning module carries more learning burden during training, and most of the existing works focus on improving the effectiveness of this module in modelling the multi-functional representations, where increasingly complex architectures are proposed~\cite{HMOE,STAR,M2M,casualint,ADIN,sass,ppnet}.
% ==
Although these works have shown promising results, these complex architectures also increase both the model complexity and the training cost, which becomes an obstacle to generalization to more business scenarios.
% ==
On the other hand, improvements for scenario-specific prediction modules are usually neglected in previous works, i.e., they only utilize simple fully-connected layers as the architecture of the predictor, which may lead to performance bottlenecks within each scenario.

% ==
In this paper, to address the above problems, we propose a novel \textbf{Opt}imizing \textbf{M}ulti-\textbf{S}cenario \textbf{M}odelling (OptMSM) framework.
% ==
We propose a novel scenario-enhanced learning module to alleviate the first problem.
% ==
Specifically, we incorporate scenario priors to partition the input feature set into scenario-specific and scenario-shared features, mapped to an embedding encoding specific information and a set of embeddings encoding shared information.
% ==
After introducing adaptive gating and orthogonality constraints on the latter to facilitate the separation of shared information corresponding to each scene, it is combined with the former to obtain the multifunctional representation.
% ==
Since neither adaptive gating nor orthogonality constraints require additional learnable parameters, and the separate modelling of feature sets eases the learning burden, the scenario-enhanced learning module provides an effective and efficient way to obtain the desired representations.
% ==
Inspired by the effectiveness of feature interactions in single-scenario modelling, we then develop a scenario-specific hypernetwork to deal with the second problem, which generates adaptive network parameters based on scenario-aware representations.
% ==
In this way, scenario-aware representations can fully interact with scenario-specific predictors to further improve performance.
% ==
Moreover, as shown in Section~\ref{cp:inter}, our framework can also be effectively integrated with existing multi-scenario models to improve performance.

\section{Related Work}\label{sec:related}
% ==
In this section, we briefly review some related works on two topics, including single-scenario modelling and multi-scenario modelling for CTR prediction.

\textbf{Single-Scenario Modeling for CTR Prediction.} 
% ==
Traditional CTR prediction aims to leverage the user interactions within a specific scenario to train an effective model for this scenario~\cite{FM,Wide_Deep,DLRM,feature}.
% ==
Most existing works on this topic focus on improving the modelling of feature interactions to enhance the performance of models, and many representative methods have been proposed.
% ==
For example, DeepFM combines factorization machine and deep network layer to model the feature interactions~\cite{DeepFM}, DCN~\cite{DCN} and DCN-V2~\cite{dcnv2} develop a novel cross-network layer to further characterize the explicit feature interactions, and APG~\cite{apg} proposes an adaptive parameter generation network for deep CTR prediction models, which can enhance the representation of feature interactions per instance with a larger parameter space.
% ==
In addition, some recent works have introduced various automated machine learning ideas to efficiently find a suitable feature interaction architecture, such as AutoFIS~\cite{AutoFis} and OptInter~\cite{OptInter}.
% ==
Overall, previous works have shown that the design of feature interaction architecture is an important factor in improving the performance of single-domain CTR models, which is neglected in multi-scenario modelling for CTR prediction.

\textbf{Multi-Scenario Modeling for CTR Prediction.}
% ==
Multi-scenario CTR modelling aims to leverage all the user interactions in different scenarios to train one or more models to serve these scenarios simultaneously~\cite{HMOE,SARNet,ADIN,STAR,casualint,sass,zeus}, where the key question is how to use shared-specific information to learn the versatile scenario-aware representations, and then use a scenario-specific architecture for per-scenario prediction.
% ==
A lot of work has been proposed to improve the effectiveness of scenario-aware representation learning.
% == 
For example, STAR~\cite{STAR} designs a novel topological dependency to fully exploit the relationship between different scenarios.
% ==
SAR-Net~\cite{SARNet} introduces a scenario-aware attention module to extract scenario-specific user features, and a corresponding gating mechanism is designed to fuse them with shared information.
% ==
CausalInt~\cite{casualint} introduces the priors on causal graphs to efficiently extract shared information and reduce negative transfer.
% ==
However, these methods will significantly increase the model parameter size and training difficulty.
% ==
Furthermore, ignoring the improvement of scenario-specific predictor architectures will lead to performance bottlenecks.

\section{Preliminary}\label{sec:preliminary}
% ==
In this section, we first give a formal definition of the multi-scenario CTR prediction task.
% ==
Given a set of scenarios $S=\{s^m\}^{M}_{m=1}$ and a set of training instance $\{(\mathbf{x},y,s^m)\}^{N}_{n=1}$, where $\mathbf{x}\in \mathbf{X}$ is the feature vector, $y\in \{0,1\}$ is the label, and $m\in \{1,\cdots, M\}$ is the scenario indicator corresponding to each instance.
% ==
The multi-scenario CTR prediction task needs to perform CTR prediction on these $M$ related scenarios,
\begin{equation}
    \hat{y} = \mathcal{F}(\mathbf{x},y,s^m),
\end{equation}
where $\mathcal{F}$ is the multi-scenarios model and $\hat{y}$ is the predicted label. 
% ==
Further, we can decompose this task into two stages, i.e., scenario-aware representation learning $f(\cdot)$ and scenario-specific prediction $g(\cdot)$,
\begin{equation}
\begin{aligned}
\label{eq:msctr}
&\mathcal{R}^{s^m}=f(\mathbf{x},s^m \mid \mathbf{W}), \\
&\hat{y} = g(\mathcal{R}^{s^m} \mid \{\mathbf{W}^{s^m}\}),
\end{aligned}
\end{equation}
where $\mathbf{W}$ and $\{\mathbf{W}^{s^m}$\} are weight parameters of the two stages, respectively. 
% ==
Therefore, the optimization objective for multi-scenario CTR prediction can be formalized as,
\begin{equation}
\label{eq:ER}
\mathcal{L}_{msm} = \sum_{n=1}^N \ell(y_n, \hat{y}_n),
\end{equation}
where $\ell$ is an arbitrary loss function, such as a cross-entropy loss.

\section{The Proposed Framework}\label{sec:method}
\begin{figure}[htbp]
    \centering
    \includegraphics[width=0.9\textwidth]{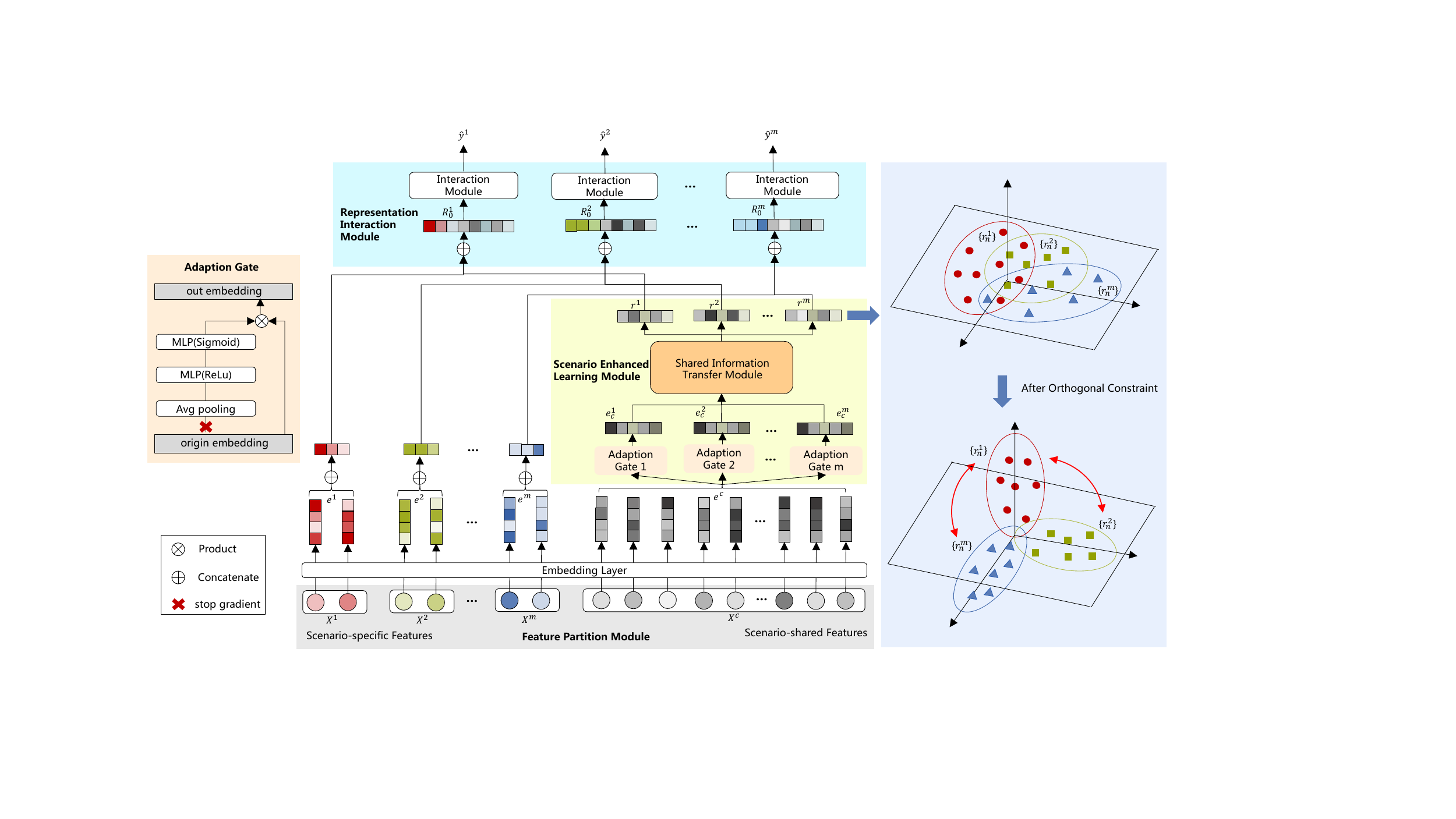}
    \caption{The architecture of our OptMSM framework.}
    \label{fig:framework}
    \vspace{-20pt}
\end{figure}

% ==
The proposed framework for optimizing multi-scenario modelling, or OptMSM for short, is shown in Fig.~\ref{fig:framework}. The OptMSM consists of three steps.
% ==
First, the input feature partition module incorporates the scenario priors to partition the input features.
% ==
Then, the scenario-enhanced learning module models the disentangled representation corresponding to each scenario from the scenario-shared features.
% ==
Finally, after combining scenario-specific information and disentangled representation, a scenario-aware representation interaction module is used to explore the interactions within each scenario to enhance predictive performance.
% ==
We will describe each module in detail based on the training process.

\subsection{The Input Features Partition Module}
% ==
To ease the model's learning burden for scenario-aware representations, we propose to divide the input features into two groups and model them separately, including scenario-specific features $\mathbf{x}^m$ and scenario-shared features $\mathbf{x}_c$, i.e. $\mathbf{x} = \{\mathbf{x}^m, \mathbf{x}_c\}$.
% ==
An example of different categories of input features is listed in Table~\ref{tab:feature}.
% ==
It can be observed that some features are specific to certain scenarios, such as \textit{scenario id}, while others are shared among all scenarios, such as \textit{gender}.
% ==
Note that previous modelling paradigms do not differentiate input feature categories. Therefore, scenario-specific features are difficult to transfer across scenarios during learning scenario-aware representations, and scenario information is hard to capture in the final prediction. Hence, the intuitive motivation for this module is to resolve these issues.
% ==
In addition, the model needs more effort to reasonably balance the modeling of two categories of features.
% ==
Next, we transform scenario-specific and scenario-shared features into corresponding low-dimensional embeddings and feed them into the following modules for further modeling.
\begin{equation}
\begin{gathered}
\label{eq:specific}
    \mathbf{e}^m = \mathbf{E}^m(\mathbf{x}^m)\quad \& \quad\mathbf{e}_c = \mathbf{E}(\mathbf{x}_c),
\end{gathered}
\end{equation}
where $\mathbf{E}^m$, $\mathbf{e}^m$, $\mathbf{E}$, $\mathbf{e}_c$ are the embedding tables and embeddings corresponding to the two category features, respectively.
\begin{table}[htbp]
    \centering
    \caption{An example of features included in the online financial recommendation platform used in the experiments.}
    \begin{tabular}{c|c}
    \toprule
    Feature Category& Example  \\
    \hline
    User Common Features & \textit{gender},\textit{age}, \textit{user behaviors}, \textit{etc.} \\
    Item Common Features & \textit{item category}, \textit{item price},\textit{ etc.} \\
    Context Common Features & \textit{time}, \textit{market condition}  \\
    User Scenario-specific Features &  \textit{user behaviors in scenarios} \\
    Item Scenario-specific Features & \textit{item statistics, item appearance in scenarios}, \textit{etc.} \\
    Context Scenario-specific Features & \textit{scenario id}, \textit{item position in scenarios}, \textit{etc.} \\
    \bottomrule
    \end{tabular}
    \label{tab:feature}
    \vspace{-20pt}
\end{table}

\subsection{The Scenario Enhanced Learning Module} \label{cp:inter}
% ==
After receiving the scenario-shared feature embeddings $\mathbf{e}_c$ generated by the previous module, we need to leverage cross-scenario information sharing and transfer to learn effective scenario-aware representations for different scenarios.
% ==
An intuitive idea is that each scenario should pay extra attention to scenario-shared features~\cite{ADIN}.
% ==
Therefore, we first introduce an adaptation gate for each scenario to refine $\mathbf{e}_c$ with scenario-specific information.
% ==
In this paper, we take Squeeze-and-Excitation (SE-Net)~\cite{SENet} as an example implementation,
\begin{equation}
\begin{gathered}
    \mathbf{z}^m = \sigma(W^m[average(\mathbf{e}_{c1}),...,average(\mathbf{e}_{ci})] + b^m),\\
    \mathbf{e}_c^m=concat([z_1^m*\mathbf{e}_{c1},...,z_i^m*\mathbf{e}_{ci}]),
\end{gathered}
\end{equation}
where $\mathbf{z}^m=\left(z_1^m,z_2^m,\cdots,z_i^m\right)$ is the refined weight vector for scenario $m$, $W^m$ and $b^m$ are the corresponding learnable parameters, and $i$ is the number of scenario-shared features.
% ==
Note that adaptive gates can be implemented differently, such as an attention layer~\cite{attention} or a perceptual layer~\cite{SARNet}, and the SE-Net block will be a lightweight approach for our purposes.
% ==
Next, a built-in shared information transfer module aims to utilize the information synergy among all the scenarios to further distinguish different concerns of different scenarios on scenario-shared information. 
% ==
The issues for this module focus on how to transfer and what to transfer. 

\textbf{How to Transfer.} 
% ==
A range of scenario-aware learning architectures have been explored in previous works on multi-scenario modeling, and they are easily integrated into this module.
% ==
Here are some examples to illustrate the process:

\begin{itemize}[topsep=0pt,noitemsep,nolistsep,leftmargin=*]
\item \textit{shared network:} The shared network aims to extract commonality from all the scenarios and can be expressed as follows,
\begin{equation}
    r_{shared}=\mathbf{MLP}_{shared}(\mathbf{e}_c^m),
\end{equation}

where $\mathbf{MLP}_{shared}$ is the multilayer perception network shared by all the scenarios. 
% ==
Note that multiple similar $\mathbf{MLP}_{shared}$ are used in MMOE, the parameters of $\mathbf{MLP}_{shared}$ are shared without explicit output in STAR, and the output of multiple $\mathbf{MLP}_{shared}$ are used as a shared expert component in PLE.

\item \textit{scenario-specific network}: The scenario-specific network aims to squeeze out scenario-specific information from shared information, in which only scenario-specific data are used,
\begin{equation}
    r_{scenario}^m=\mathbf{MLP}^m(\mathbf{e}_c^m),
\end{equation}
where $\mathbf{MLP}^m$ is the scenario-specific network. 
% ==
Note that the number of $\mathbf{MLP}^m$ can be set according to the plugged module, e.g., 0 in MMOE, equal to the number of scenarios in STAR, and a predefined value in PLE.

\item \textit{transferring layer}: In some previous works, different methods are introduced to jointly model the above two networks. 
% ==
For example, STAR proposes the FCN topology dependence, and PLE introduces the gated network. 
% ==
To illustrate the transfer process, we use FCN as an example,
\begin{equation}
    r_{transfer}^m=FCN(\mathbf{e}_c^m) = (W_{shared} \otimes W^m) \cdot \mathbf{e}_c^m + b_{shared} + b^m,
\end{equation}
where $\{W_{shared}, b_{shared}\}$ and $\{W^m,b^m\}$ are parameters in $\mathbf{MLP}_{shared}$ and $\mathbf{MLP}^m$, respectively, and $\otimes$ denotes element-wise multiplication.
\end{itemize}
Finally, this module will generate representations for all the scenarios, denoted as $\{r^m \mid m \in [1, M]\}$.

\textbf{What to Transfer.} Note that the scenario-aware representations are learned based on the model that mixes samples from all the scenarios. 
% ==
As a result, negative transfer often occurs, which perturbs the scenario-aware representations and misleads subsequent top-level predictions.
% ==
A critical issue to mitigate the negative transfer effect is disentangling the representations between different scenarios.
% ==
Inspired by the disentangled representation learning~\cite{orth}, we propose an explicit orthogonality constraint on the representation obtained above as an auxiliary loss to achieve this goal. 
% ==
Note that the number of samples in all the scenarios is usually unbalanced, and it is difficult to deal with the constraints of cross-sample representations.
% ==
Therefore, we propose a strategy for enhanced learning.
% ==
More specifically, for a sample $b$, we generate its representations in all the scenarios, i.e., $\{r_b^1, \cdots, r_b^m\}$.
% ==
Only one representation corresponding to the real scenario will be used for prediction in subsequent layers, while the others are used as \textit{contrastive representations} to compute the orthogonality constraint.
% ==
Orthogonal constraints will make these representations perpendicular to each other to ensure independence and successfully disentangle scenario-specific information.
% ==
Note that the idea behind this strategy is similar to contrastive loss~\cite{CL}. 
% ==
Formally, the loss can be expressed as follows,
\begin{align}
\label{eq:orth}
    \mathcal{L}_{orth} = \sum_{\substack{i\neq j \\ b\in B}}<r_{b}^i, r_{b}^j> \text{,} \quad
    <r_{b}^i, r_{b}^j> = \frac{r_{b}^i \cdot r_{b}^j}{\Arrowvert r_{b}^i\Arrowvert_2\cdot \Arrowvert r_{b}^j\Arrowvert_2}
\end{align}
where $\Arrowvert\cdot\Arrowvert_2$ refers to the $\mathit{l}_2$ norm, and $B$ is the size of the mini-batch. 
% ==
Note that although the loss is conducted on $C_m^2$ pairs, it can be efficiently implemented in a vectorized manner at the mini-batch level and avoids loops.

\subsection{The Scenario-Aware Representation Interaction Module}
\begin{figure}[!htbp]
    \centering
    \includegraphics[width=0.5\textwidth]{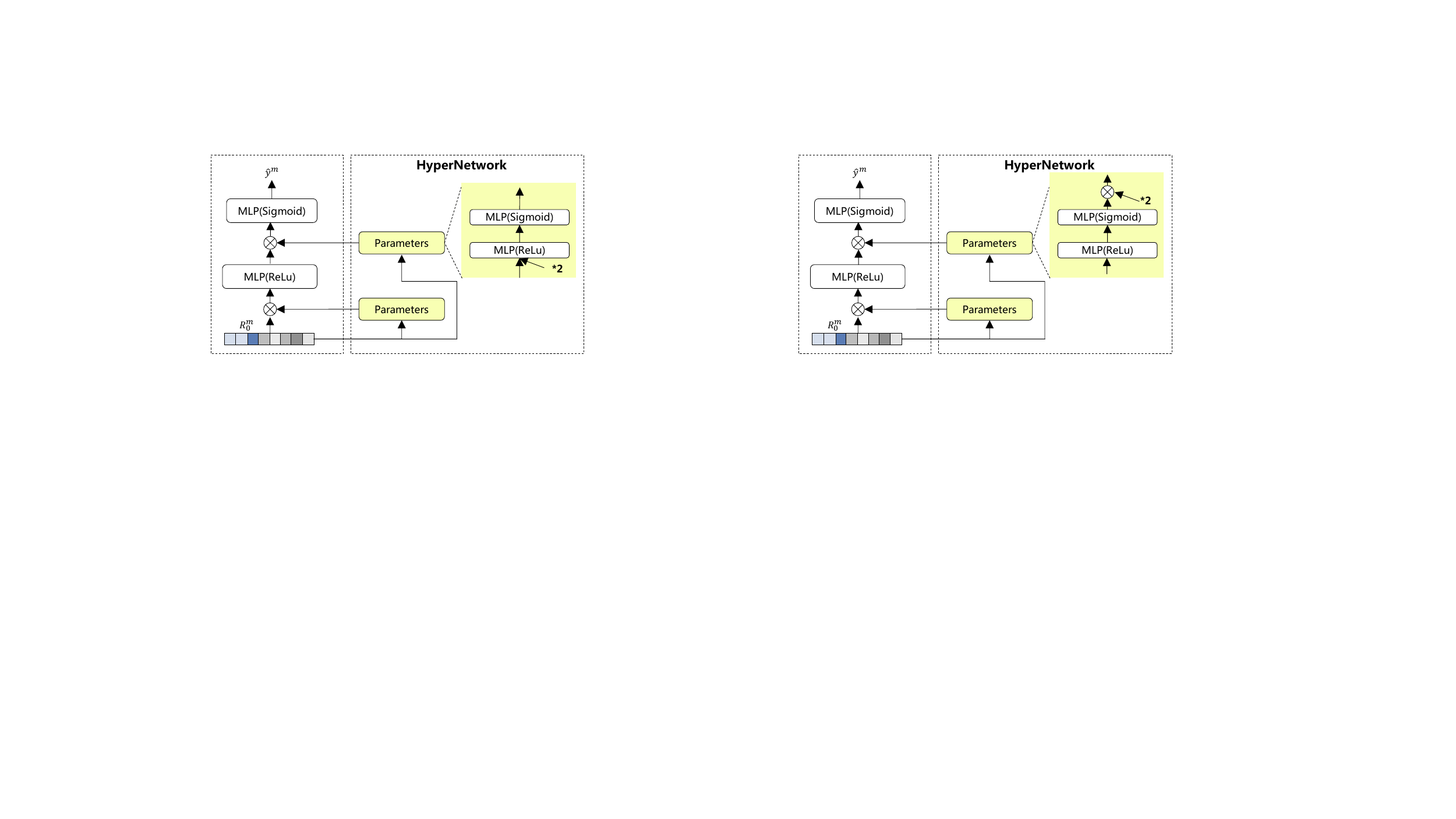}
    \caption{The scenario-aware hypernetwork for parameters generation.}
    \label{fig:hyper}
\end{figure}

% ==
Although we get the disentangled scenario-aware representation, we still need to augment the representation with prior scenario-specific features in Eq.\eqref{eq:specific}. 
% ==
On the one hand, scenario-specific information has a solid induction to the corresponding scenario, which helps the final prediction.
% ==
On the other hand, considering complex interactions has been shown to benefit the performance of single-scenario CTR modeling.
% ==
Therefore, to give the prediction more perception of prior information, we design a hypernetwork adaptively generating scenario-aware parameters~\cite{ppnet,hypernetwork}, which provides a full representation interaction.
% ==
We give a detailed illustration of this module as shown in Fig.~\ref{fig:hyper}.

% ==
To preserve the priors, we only concatenate the prior scenario-specific features embeddings $\mathbf{e}^m$ with disentangled scenario-aware representation $r^m$,
\begin{equation}
    \mathcal{R}_0^m = r^m \oplus \mathbf{e}^m.
\end{equation}
% ==
We then adopt a two-layer perception to generate parameters from the representations, i.e.,
\begin{equation}
\begin{gathered}
    \mathbf{\mathbb{R}}_{0l} = Relu(w_0\mathcal{R}_0^m + b_0),\\
    \label{eq:rep}
    \mathbf{\mathbb{R}}_{1l} = 2\star \sigma(w_1 \mathbf{\mathbb{R}}_{0l} + b_1),
\end{gathered}
\end{equation}
where $\sigma$ is sigmoid function, $\mathbf{\mathbb{R}}_{1l}$ has the same shape as $\mathcal{R}_l^m$, and $l$ is is the current layer number. 
% ==
Setting the coefficient to 2 in Eq.\eqref{eq:rep} is to scale the mean of sigmoid output to 1. 
% ==
After parameters are generated, we interact $\mathbf{\mathbb{R}}_{1l}$ with each layer in each scenario-specific predictor,
\begin{equation}
    \mathcal{R}_l^m = \mathcal{R}_l^m \otimes \mathbf{\mathbb{R}}_{1l}, \ l \in \{0, \cdots, L-1\}
\end{equation}
where $\mathcal{R}_l^m$ is the latent output of layer $l$ in the scenario $m$, and $L-1$ is the number of layers in each scenario. Finally, the final score for $m$-th scenario can be get,
\begin{equation}
\label{eq:pred}
    \hat{y}^m = \sigma(W_{L-1}R_{L-1}^m+b_{L-1}),
\end{equation}
where $W_{n-1},b_{n-1}$ is the parameters of classifier. After combining Eq.\eqref{eq:ER}, \eqref{eq:orth} and \eqref{eq:pred}, We can get the final optimization objective,
\begin{equation}
    \mathcal{L} = \mathcal{L}_{msm} + \lambda \cdot \mathcal{L}_{orth},
\end{equation}
where $\lambda$ is a hyper-parameter controlling the orthogonality constraint. 

\section{Experiments}
\label{sec:exp}
In this section, we conduct comprehensive experiments with the aim of answering the following five key questions. 
\begin{itemize}[topsep=0pt, noitemsep, nolistsep, leftmargin=*]
    \item \textbf{RQ1}: Could OptMSM achieve superior performance compared with mainstream multi-scenario models?
    \item \textbf{RQ2}: Could OptMSM transfer to more multi-scenario models?
    \item \textbf{RQ3}: How does each module of OptMSM contribute to the final results?
    \item \textbf{RQ4}: Does OptMSM really get the optimal scenario-aware representation?
    \item \textbf{RQ5}: How does OptMSM perform in real-world recommendation scenarios?
\end{itemize}

\subsection{Experiment Setup}
\textbf{Datasets.}
% ==
We conduct our offline experiments on three datasets, including two publicly multi-scene CTR benchmark datasets (Ali-CCP and AliExpress) and a private product dataset.  
% ==
Ali-CCP\footnote{https://tianchi.aliyun.com/dataset/408} is collected from the traffic log of Tabao, and we divide logs into three scenarios according to the \textit{scenario id}. 
% ==
AliExpress\footnote{https://tianchi.aliyun.com/dataset/74690} is collected from the AliExpress search system, which contains user behaviours from five countries. 
% ==
We consider each country as an advertising scenario and select four countries in our experiments following the setting of previous work~\cite{AESM2}. 
% ==
The real product dataset comes from the financial business scenario of Tencent Licaitong, and we collect consecutive 4 weeks of user feedback logs from four scenarios, respectively. 
% ==
For Ali-CCP, following previous work~\cite{AITM}, we use all the single-valued categorical features and take 10\% of the train set as the validation set to verify models.
% ==
For AliExpress, we split the training set and test set according to the settings in the original paper~\cite{HMOE}. 
% ==
For the production dataset, we keep data on the last day as the test set, and the rest as the training and validation sets.
% ==
Table~\ref{tab:stat} summarize the statistics for these datasets. 
% ==
We can observe that the data distribution in Ali-CCP and Production is obviously unbalanced.

\begin{table}[htbp]
\centering
\caption{Statistics of datasets used in offline experiments. For impression and click, the percentages in each scenario are given in brackets.}
\begin{tabular}{c|c|c|c}
\toprule
 Dataset& \#Scenarios &\#Impression & \#Click \\ 
 \hline
 Ali-CCP& 3 & \makecell[c]{85,316,519\\(0.75/37.79/61.46)} & \makecell[c]{3,317,703\\(0.84/38.91/60.25)}   \\ \hline
 AliExpress& 4 & \makecell[c]{103,814,836\\(17.07/26.04/30.51/26.38)} &\makecell[c]{2,215,494\\(17.02/24.49/38.15/20.34)}   \\ \hline
 Production & 4& \makecell[c]{823,972,400\\(68.96/3.93/8.05/19.06)} & \makecell[c]{59,466,088\\(47.38/7.21/14.09/31.32))}\\
 \bottomrule
\end{tabular}
\label{tab:stat}
\end{table}

\textbf{Comparison Models.} 
To verify the effectiveness of our proposed framework, we compare OptMSM with the following models. \textbf{Mix}: The model with a 3-layer fully-connected network is trained with a mixture of samples from all Scenarios; \textbf{S-B}: We share the embedding table across scenarios, and each scenario-specific network is the same as Mix, i.e., \textbf{s}hared \textbf{b}ottom model; \textbf{MMoE}~\cite{MMOE}: We adopt a shared Mixture-of-Experts model, where each expert is a 3-layer fully-connected network and the number of experts equals $2*\#scenarios$; \textbf{HMOE}~\cite{HMOE}: Except for explicit relatedness in the label space introduced by HMOE, the other settings are the same as MMOE; \textbf{PLE}~\cite{ple}: The core module of PLE is CGC (Customized Gate Control), which consists of scenario-specific experts and shared experts. We keep the number of the former the same as MMOE with two additional shared experts; \textbf{STAR}~\cite{STAR}: This model consists of a centered network shared by all scenarios and the scenario-specific network for each scenario. The architectures of all networks are the same as Mix; and \textbf{PEPNet}~\cite{ppnet}: This model adopts personalized prior information to enhance embedding and parameter personalization, and only has scenario-specific towers for predictions.

\textbf{Implementation Details \& Evaluation Settings.}
All models are implemented on Tensorflow~\cite{tf} and trained with Adam optimizer~\cite{adam}. We tune learning rate from [$10^{-2}$, $10^{-3}$, $10^{-4}$, $10^{-5}$], \textit{L}2 weight from [$10^{-3}$, $10^{-4}$, $10^{-5}$, $10^{-6}$], and dropout rate from [0.1, 0.2, 0.3, 0.4]. The batch sizes for each dataset are set as 2048, 2048, and 512, respectively. The embedding dimensions are set as 20, 10, and 10. Besides, the hidden layers of the fully connected network are fixed to [256, 128, 32]. 
Following the previous works~\cite{DeepFM,STAR}, we use two common metrics in CTR prediction, i.e., AUC (Area Under ROC) and Logloss (based on cross-entropy). 

\subsection{RQ1: Overall Performance}
We show the overall performance of our OptMSM and other baselines in Table~\ref{tab:performance}. We summarize our observations below: 1) OptMSM generally outperforms baselines in most scenarios in three datasets. Specifically, OptMSM performs consistently well in three scenarios in the Ali-CCP dataset and improves significantly in the first sparse scenario. In the other two datasets, our OptMSM performs better in most scenarios to different degrees. Although OptMSM achieves the second performance in some scenarios, note that the difference is within 0.1\%, which is also acceptable considering OptMSM gains statistical improvements in other scenarios; 2) On the whole, MSM can boost performance in all scenarios compared with the model trained with mixed data. However, this model performs slightly better than others in scenarios with sparse training samples. For example, the Mix model performs better in Ali-CCP S1 and Production S2. The possible reason for this is that samples in other scenarios are far more than these two scenarios and can directly help prediction in these two scenarios; and 3) PEPNet performs consistently better in AliExpress compared with other baselines while achieving relatively poor performance in other skewed datasets. Note that the distribution of AliExpress is more balanced than the other two datasets. Hence, this comparison directly verifies the effectiveness of information priors in some datasets and indirectly reflects that positive transfer is important when spares scenarios exist.

\begin{table}[htbp]
    \centering
    \caption{The overall performance over three datasets. The boldface and underline indicate the highest score of all the models and baselines. $\star$ indicates significant level p-value $<0.05$.}
    
    \resizebox{.95\textwidth}{!}{
    \begin{tabular}{c|c|c|c|c|c|c|c|c|c|c}
    \toprule
         & Scenario & Metric & Mix & S-B & MMOE & HMOE & PLE & STAR & PepNet & OptMSM\\
         \hline
         \multirow{6}*{\rotatebox{90}{Ali-CCP}}&\multirow{2}*{S1}& AUC&0.5921&0.5899&0.5955&\underline{0.5979}&0.5943&0.5924&0.5941&$\mathbf{0.6023}^{\star}$\\
         &&Logloss&0.1838&0.1855&0.1811&\underline{0.1801}&0.1811&0.1906&0.1922&$\mathbf{0.1782}^{\star}$\\
         \cline{2-11}
         &\multirow{2}*{S2}&AUC&0.6166&0.6202&0.6183&0.6214&0.6198&\underline{0.6246}&0.6203&$\mathbf{0.6257}^\star$\\
         &&Logloss&0.1673&0.1663&0.1657&0.1662&\underline{0.1657}&0.1715&0.1724&$\mathbf{0.1648}^\star$ \\
         \cline{2-11}
         &\multirow{2}*{S3}&AUC&0.6141&0.6164&0.6151&\underline{0.6183}&0.6165&0.6175&0.6168&$\mathbf{0.6231}^\star$ \\
         &&Logloss&0.1641&0.1600&0.1596&0.1601&\underline{0.1596}&0.1601&0.1693&$\mathbf{0.1587}^\star$\\
         \hline
         \multirow{8}*{\rotatebox{90}{AliExpress}} & \multirow{2}*{NL}&AUC&0.7256&0.7253&0.7257&\underline{0.7261}&0.7256&0.7257&0.7258&$\mathbf{0.7286}^\star$ \\
         &&Logloss&0.1087&0.1086&0.1081&0.1080&0.1079&0.1084&\underline{0.1078}&$\mathbf{0.1077}$ \\
         \cline{2-11}
         &\multirow{2}*{FR}&AUC&0.7247&0.7256&0.7258&0.7260&0.7263&0.7258&\underline{$\mathbf{0.7266}$}&0.7256 \\
         &&Logloss&0.1010&0.1013&0.1009&0.1007&0.1008&0.1009&\underline{0.1006}&$\mathbf{0.1004}$ \\
         \cline{2-11}
         &\multirow{2}*{ES}&AUC&0.7272&0.7276&0.7281&0.7285&0.7279&0.7277&\underline{0.7290}&$\mathbf{0.7301}^\star$ \\
         &&Logloss&0.1211&0.1210&0.1207&0.1207&0.1208&0.1211&\underline{0.1204}&$\mathbf{0.1201}$ \\
         \cline{2-11}
         &\multirow{2}*{US}&AUC&0.7084&0.7059&0.7082&0.7084&0.7084&0.7073&\underline{0.7088}&$\mathbf{0.7108}^\star$ \\
         &&Logloss&0.1015&0.1008&0.1008&0.1006&0.1006&0.1007&\underline{$\mathbf{0.1004}$}&0.1005 \\
         \hline
         \multirow{8}*{\rotatebox{90}{Production}}&\multirow{2}*{S1}&AUC&0.8718&0.8853&0.8866&0.8811&0.8872&\underline{0.8875}&0.8866&$\mathbf{0.8890}^\star$ \\
         &&Logloss&0.0951&0.0914&0.0954&0.0982&0.0958&\underline{0.0862}&0.0956&$\mathbf{0.0848}^\star$\\
         \cline{2-11}
         &\multirow{2}*{S2}&AUC&0.8997&0.9065&0.9069&0.9004&$\mathbf
{\underline{0.9077}}$&0.9069&0.9068&0.9071 \\
         &&Logloss&\underline{$\mathbf{0.0246}$}&0.0248&0.0256&0.0259&0.0258&0.0259&0.0317&0.0247 \\
         \cline{2-11}
         &\multirow{2}*{S3}&AUC&0.8414&0.8478&0.8491&0.8502&0.8496&\underline{0.8515}&0.8507&$\mathbf{0.8524}^\star$ \\
         &&Logloss&0.0361&0.0288&0.0286&0.0286&0.0288&\underline{0.0276}&0.0340&$\mathbf{0.0273}$ \\
         \cline{2-11}
         &\multirow{2}*{S4}&AUC&0.8665&0.8765&0.8759&\underline{0.8774}&0.8768&0.8756&0.8773&$\mathbf{0.8808}^\star$ \\
         &&Logloss&0.0569&0.0584&0.0581&0.0586&0.0585&\underline{0.0575}&0.0654&$\mathbf{0.0538}^\star$ \\
         \bottomrule
    \end{tabular}
    }
    \label{tab:performance}
    \vspace{-20pt}
\end{table}

\subsection{RQ2: Transferability Analysis}

\begin{figure}[htbp]
    \centering
    \subfigure{
        \includegraphics[width=0.3\textwidth]{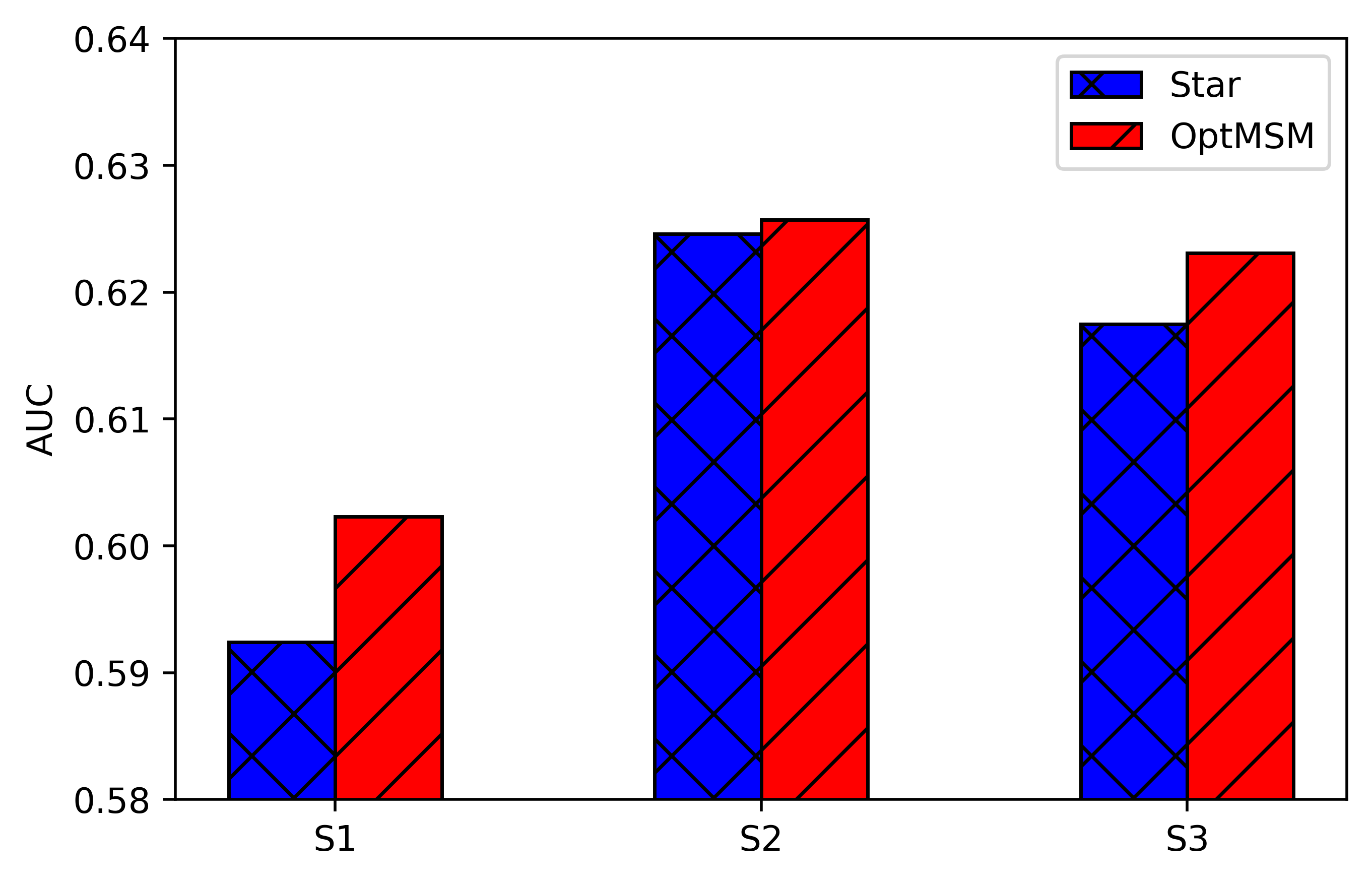}
    }
    \subfigure{
        \includegraphics[width=0.3\textwidth]{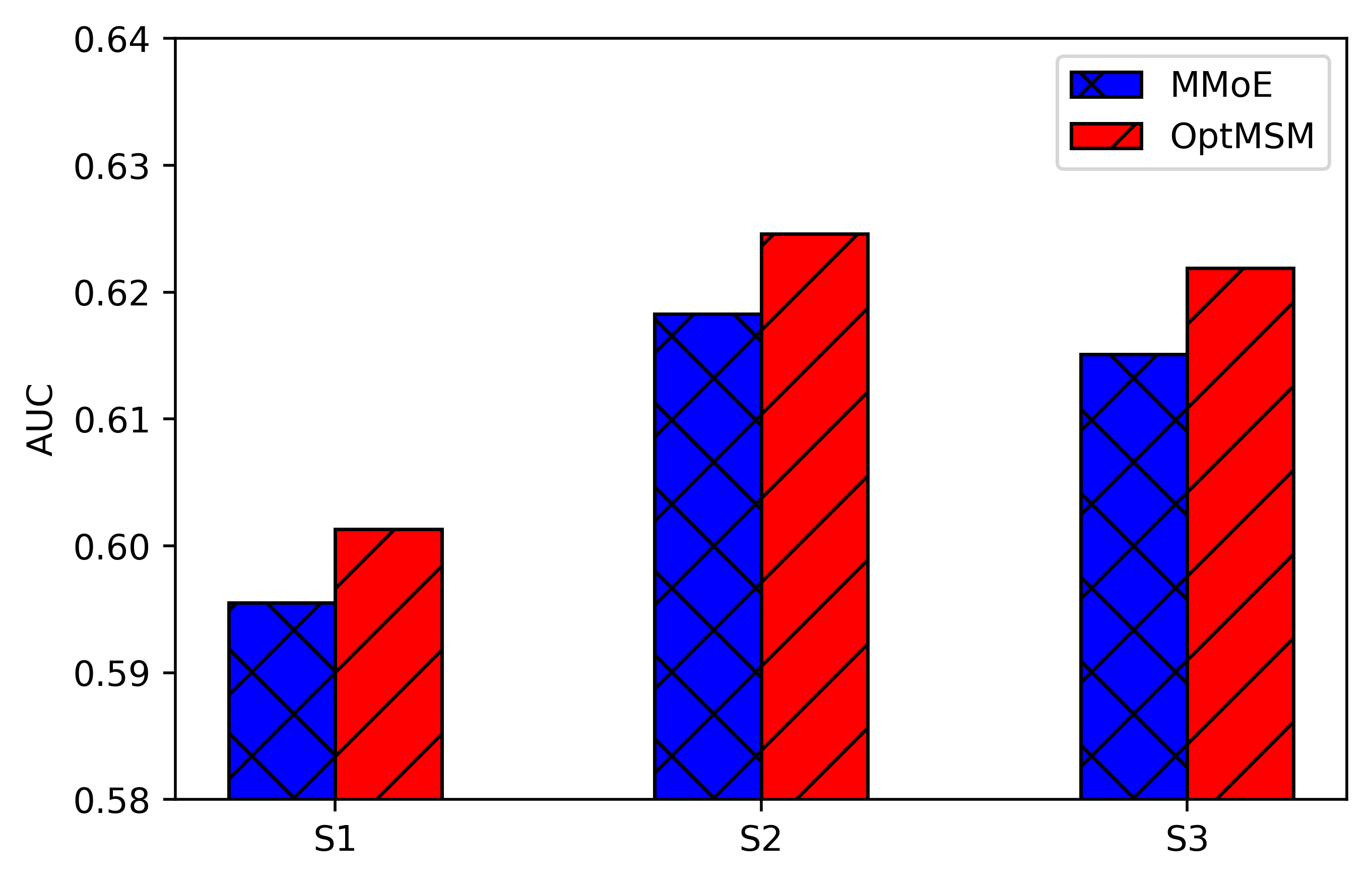}
    }
    \subfigure{
        \includegraphics[width=0.3\textwidth]{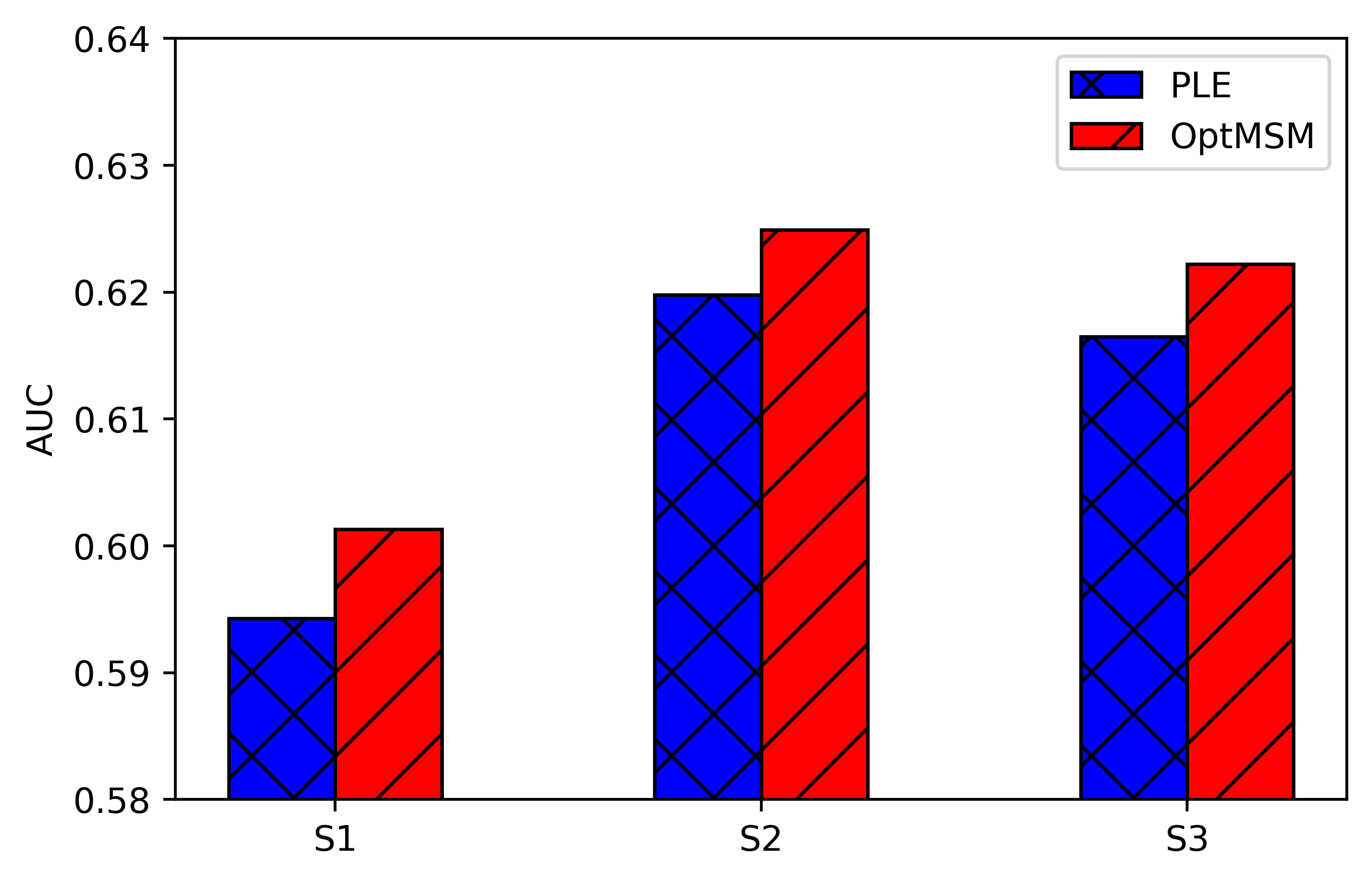}
    }
    \caption{Transferable analysis of OptMSM with different operation on Ali-CCP.}
    \label{fig:ccp_arch}
    \vspace{-20pt}
\end{figure}

In this subsection, we investigate the transferability of our framework. We introduce FCN as our shared information transfer module in our framework. Here we extend our framework to other operation modules to illustrate whether our framework really optimizes the key factors for these modules. As shown in Fig.~\ref{fig:ccp_arch}, we extend our OptMSM with transfer operations, including FCN, MoE, and CGC. Compared with the corresponding model, OptMSM improves its performance in all the scenarios, further validating the effectiveness of our design optimizing module. To investigate whether the additional optimization will bring a lot of computation cost, we report the training time of these models in Table~\ref{tab:cost}. Notably, the increment of training time of OptMSM is acceptable.

\begin{table}[htbp]
    \centering
    \caption{Training cost comparison on the Ali-CCP.}
    \begin{tabular}{c|c|c|c|c|c|c}
    \toprule
         Model& Star&OptMSM(FCN) & MMoE&OptMSM(MoE) & PLE & OptMSM(CGC)  \\
         \hline
         Cost (s)& 684 & 716 (+4.68\%) & 692 & 724 (+4.62\%)& 941&982 (+4.36\%)\\
    \bottomrule
    \end{tabular}
    \label{tab:cost}
    \vspace{-20pt}
\end{table}

\subsection{RQ3\&RQ4: Ablation Study}
\begin{figure}
   \centering
   \includegraphics[width=0.5\textwidth]{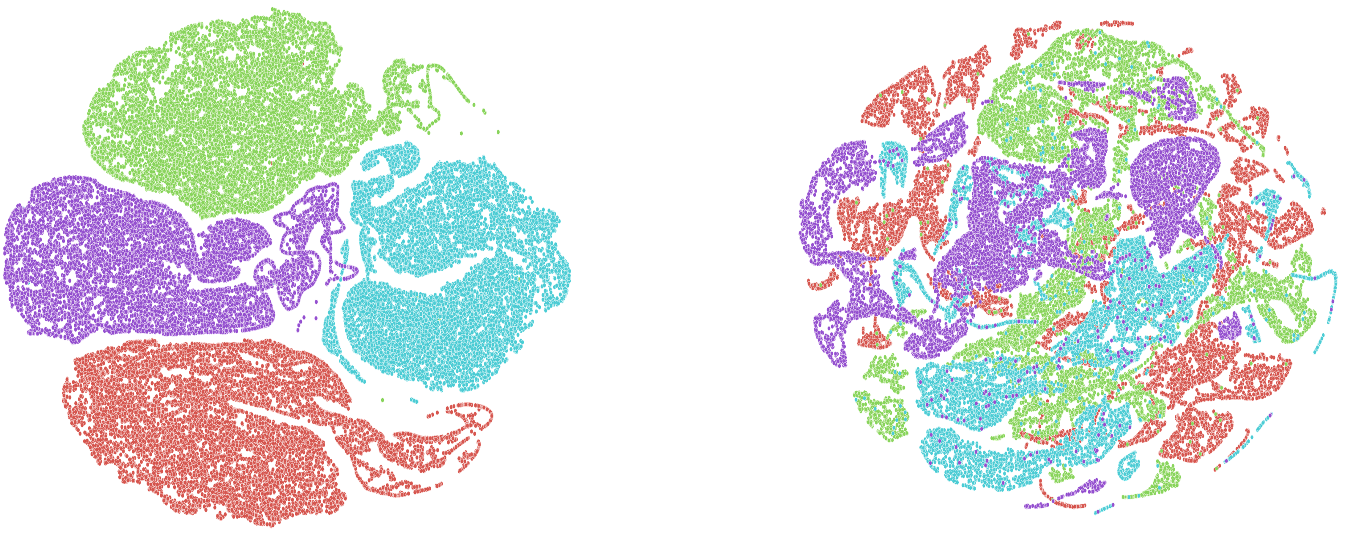}
   \caption{Visualization results on the representation in AliExpress. Left: with orthogonal constraint; Right: without orthogonal constraint.}
   \label{fig:representation}
    \vspace{-10pt}
\end{figure}

In this subsection, we validate the contribution of each component of OptMSM. We conduct a series of ablation studies over the datasets by examining the AUC after removing each component. The results are summarized in Table~\ref{tab:ablation_ccp} and~\ref{tab:ablation_express}. The observations are summarized as follows: (1) All three components play important roles in optimizing different architectures, proving our optimizing framework's effectiveness. (2) In both datasets, removing orthogonal constraints generally suffers from the most decrement in AUC, which means the disentangled representation is effective. (3)Because of the significant improvement of PEPNet in AliExpress, removing hypernetwork in AliExpress is harmful to our framework, which indicates that our framework optimizing scenario-specific prediction module is useful. As the disentangled representation is a key factor in our OptMSM, we further illustrate visual results by comparing the t-SNE~\cite{t-sne} representations with and without orthogonal constraint in Fig.~\ref{fig:representation}. Note that our constraint is effective in explicitly disentangling representation.

\begin{table}[htbp]
    \centering
    \caption{Ablation study on OptMSM with FCN for Ali-CCP. w/o means removing the corresponding component, and the relative decrement is reported in the brackets.}
    \begin{tabular}{c|c|c|c}
    \toprule
         Model& S1 & S2 & S3 \\
         \hline
         OptMSM& 0.6023&0.6257&0.6231 \\
         \hline
         \hline
         w/o priors& 0.6014 (-0.15\%) & 0.6247 (-0.16\%)&0.6222 (-0.14\%)\\
         w/o constraint& 0.6010 (-0.22\%)& 0.6246 (-0.18\%)& 0.6219 (-0.19\%)\\
         w/o hypernetwork & 0.6016 (-0.12\%)&0.6249 (-0.13)&0.6223 (-0.13\%) \\
         \bottomrule
    \end{tabular}
    \label{tab:ablation_ccp}
    \vspace{-5pt}
\end{table}
\begin{table}[htbp]
    \centering
    \caption{Ablation study on OptMSM with CGC for AliExpress. w/o means removing the corresponding component, and the relative decrement is reported in the brackets. }
    \begin{tabular}{c|c|c|c|c}
    \toprule
       Model  & NL & FR & ES & US \\
       \hline
        OptMSM & 0.7290 & 0.7268& 0.7312 & 0.7117 \\
        \hline
        \hline
        w/o priors& 0.7288 (-0.03\%) & 0.7260 (-0.11\%)&0.7302 (-0.14\%) & 0.7112 (-0.07\%)\\
         w/o constraint& 0.7277 (-0.18\%)& 0.7263 (-0.07\%)& 0.7301 (-0.15\%)&0.7108 (-0.13\%)\\
         w/o hypernetwork & 0.7280 (-0.14\%) &0.7265 (-0.04)&0.7302 (-0.14\%)& 0.7107 (-0.14\%) \\
         \bottomrule
    \end{tabular}
    \label{tab:ablation_express}
\end{table}

% \begin{figure}[htbp]
% \centering
% \begin{minipage}[t]{0.483\textwidth}
% \centering
% \includegraphics[width=0.8\textwidth]{pic/representation.png}
% \caption{Visualization on the representation in AliExpress with and without orthogonal constraint.}
% \label{fig:representation}
% \end{minipage}
% \hspace{2pt}
% \begin{minipage}[t]{0.483\textwidth}
% \centering
% \includegraphics[width=0.8\textwidth]{pic/online.pdf}
% \caption{Overview of the financial product recommender system, including online service and offline training.}
% \label{fig:online}
% \end{minipage}
% \end{figure}

\subsection{RQ5: Online Experiments}\label{sec:online}
\begin{figure}[htbp]
   \vspace{-15pt}
   \centering
   \includegraphics[width=0.5\textwidth]{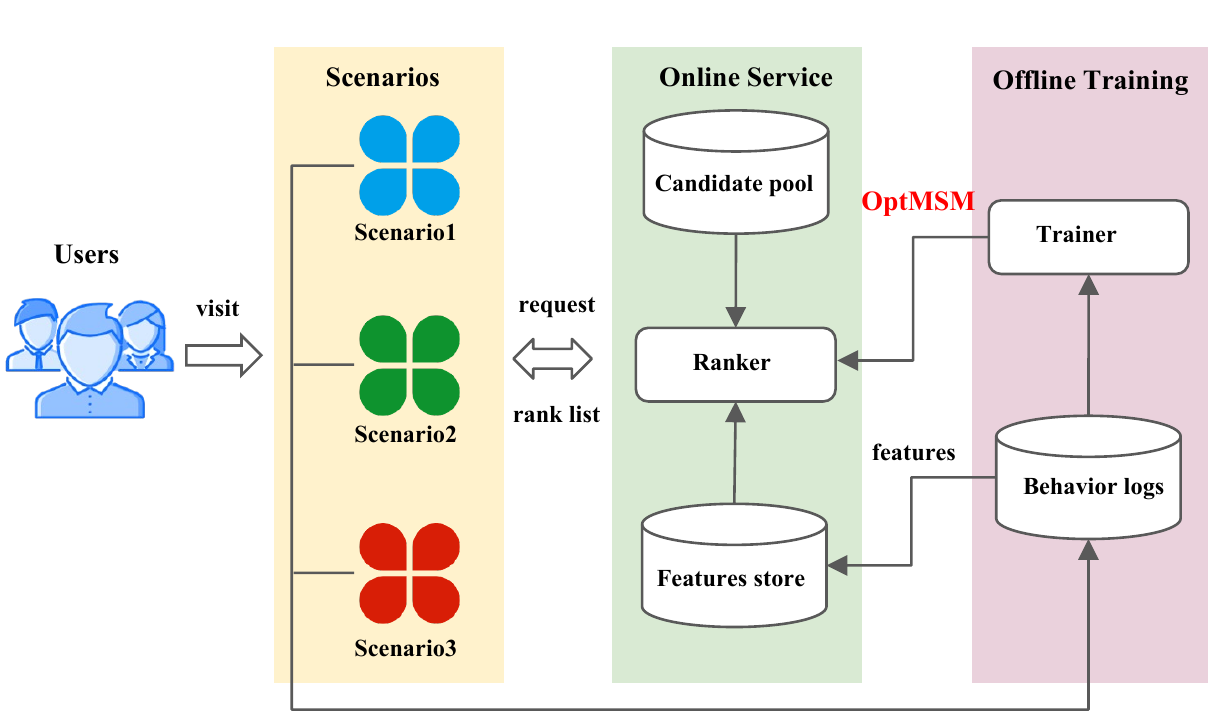}
   \caption{Overview of the financial product recommender system.}
   \label{fig:online}
    \vspace{-10pt}
\end{figure}
\begin{figure}[htbp]
    \vspace{-10pt}
    \centering
    \subfigure[Scenario 1]{
        \includegraphics[width=0.45\textwidth]{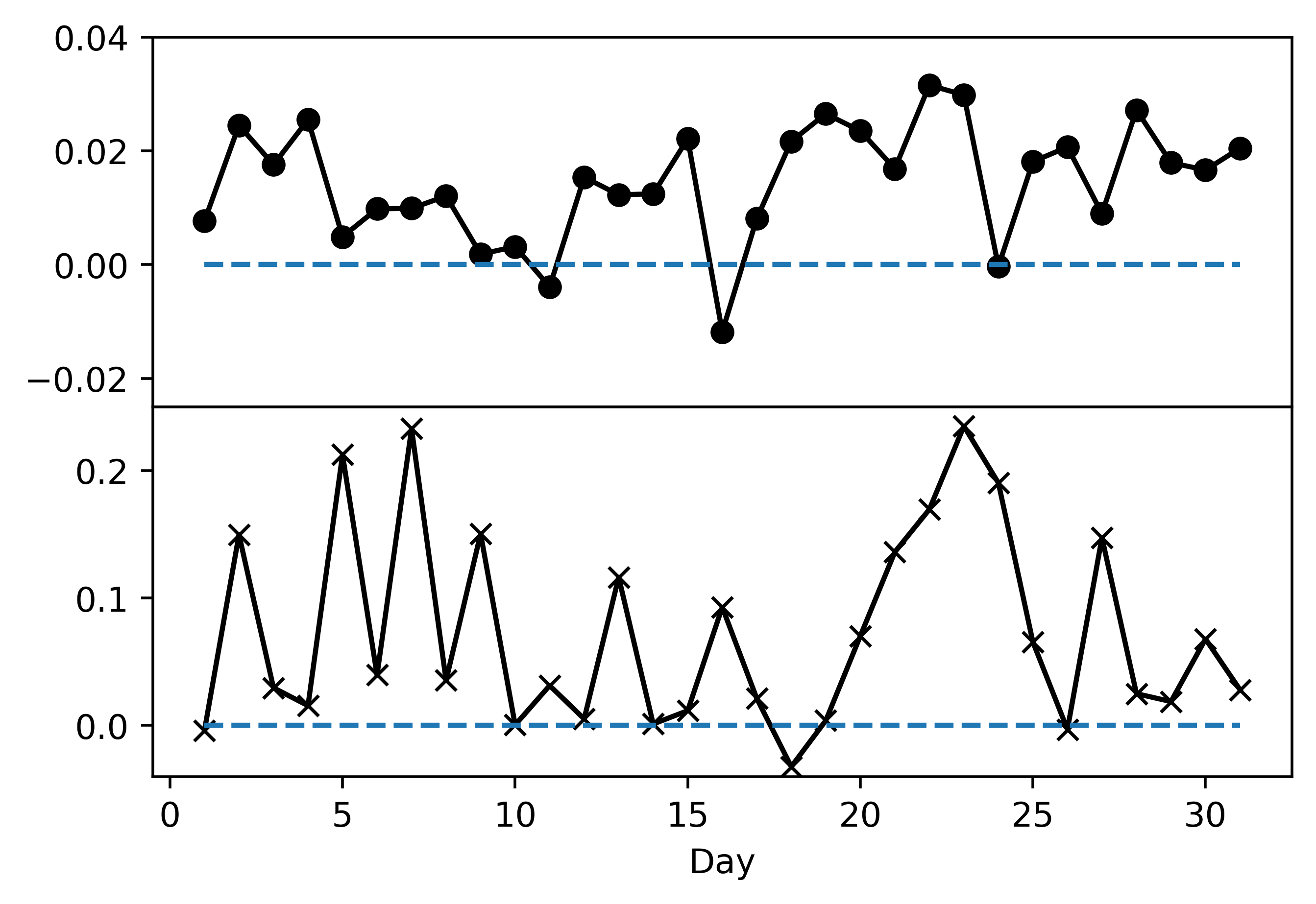}
    }
    \subfigure[Scenario 2]{
        \includegraphics[width=0.45\textwidth]{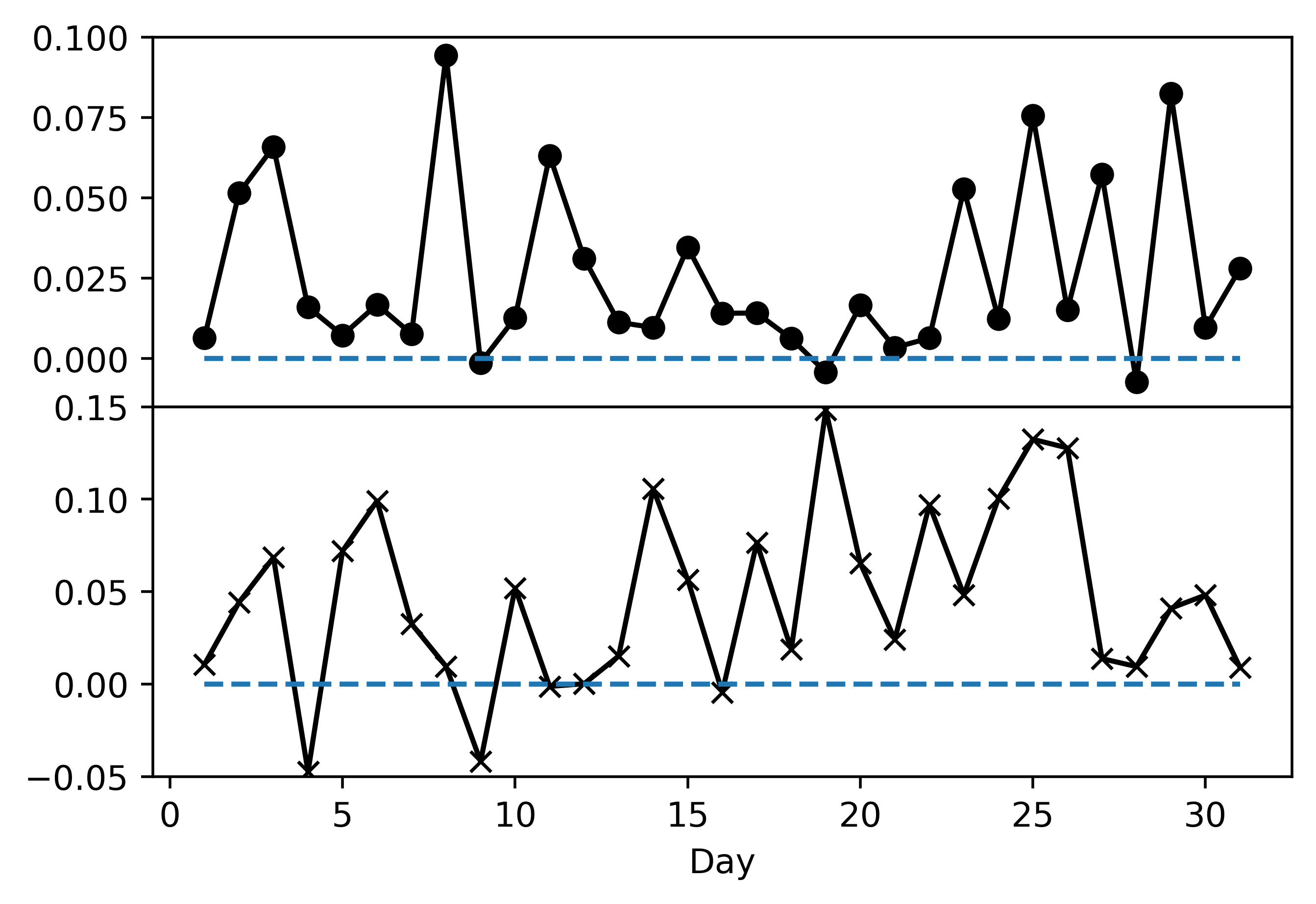}
    }
    \quad
    \subfigure[Scenario 3]{
        \includegraphics[width=0.45\textwidth]{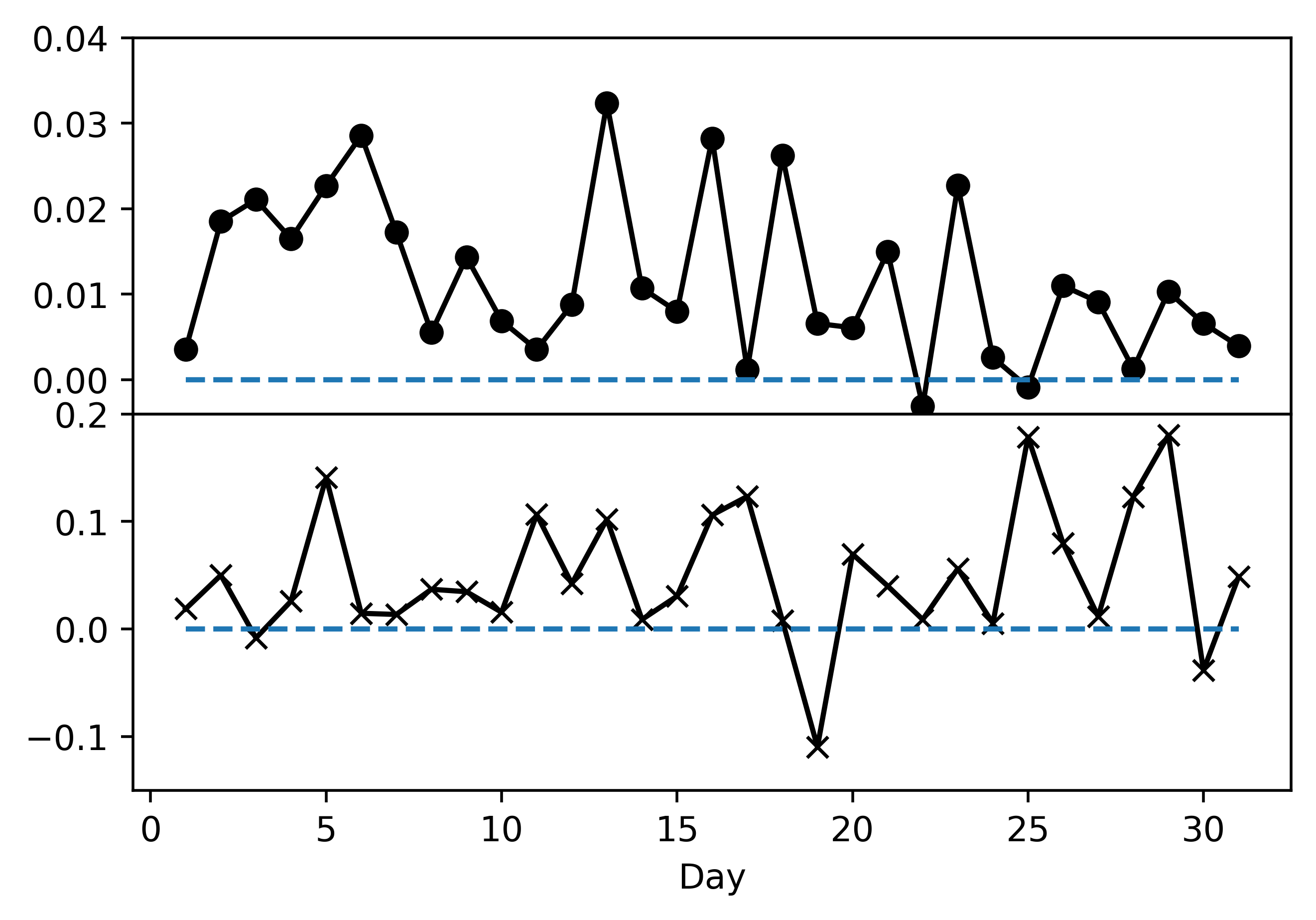}
    }
    \subfigure[Scenario 4]{
        \includegraphics[width=0.45\textwidth]{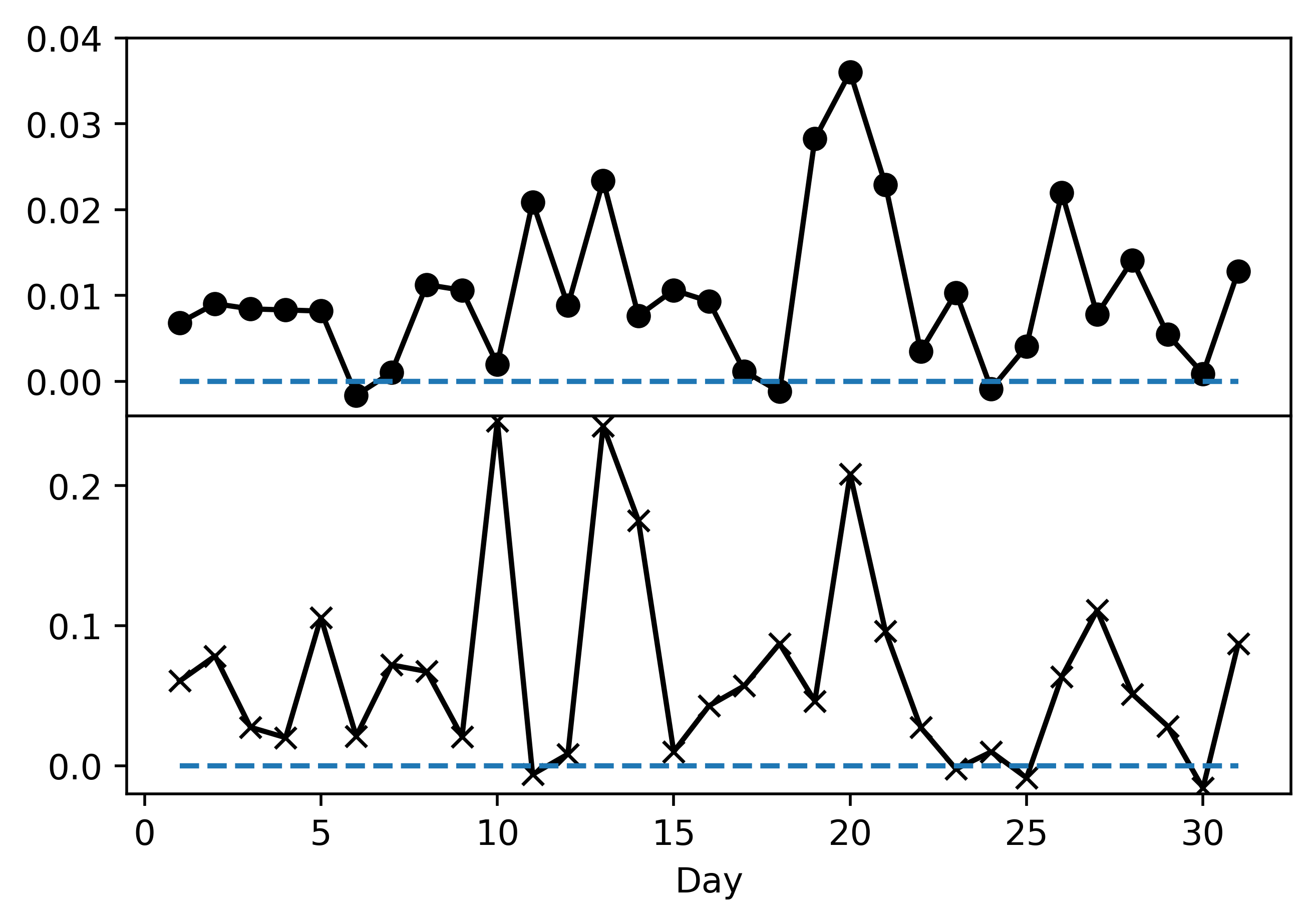}
    }
    \caption{Online relative improvement ratios in four scenarios in consecutive four weeks. (Upper is the CTR improvement, Bottom is the PAPM improvement).}
    \label{fig:online_result}
    \vspace{-20pt}
\end{figure}

In this subsection, we report the online experiment results of our OptMSM in a financial product recommender system for four consecutive weeks, and the results further verify the effectiveness of our OptMSM.
% ==
Firstly, we briefly present the recommender system overview, shown in Fig.~\ref{fig:online}. This system has two main components: Online Service and Offline Training respectively. When users access any scenarios, a rank list request will be sent to the online service. Meanwhile, the user's attributes and contextual features will also be sent to the ranker, which utilizes the offline model to predict the score. In offline training, the ranker leverages behaviour historical logs, and the trainer trains the model based on the logs daily. Our OptMSM trains a unified model here to serve multiple scenarios.
% == 
We deploy the OptMSM on four scenarios in this financial product recommender platform, which serves millions of daily active users. And the model is trained in a single cluster, where each node contains 96-core Intel(R) Platinum 8255C CPU, 256GB RAM, and 8 NVIDIA TESLA A100 GPU cards. Besides using Click Through Rate (CTR)(i.e. $\frac{\#click}{\#impression}$), a commonly-used online evaluation metric, we also use purchase amount per mille (PAPM), defined as $\frac{\#purchase\_amount}{\#impression}\times 1000$. Briefly, our OptMSM improve the overall performance, achieving \textbf{+1.42\%}, \textbf{+1.76\%}, \textbf{+1.26\%} and \textbf{+0.84\%} lift on CTR, and \textbf{+6.58\%},\textbf{+7.10\%},\textbf{+5.82\%} and \textbf{+6.90\%} lift on PAPM over 4 scenarios. The daily improvements are illustrated in Fig.~\ref{fig:online_result}.

\section{Conclusion}\label{sec:conclusion}
In this paper, we propose a framework named OptMSM, which can optimize multi-scenario modeling with disentangled representation and scenario-specific interaction. First, we partition input features into two separate feature sets incorporating scenario priors, including scenario-specific and scenario-shared features. Then we design a scenario-enhanced learning module with plugged scenario-shared information transfer. With orthogonal constraints on both scenario-aware representation and contrastive representations, we obtain the disentangled representation. Finally, the scenario-specific interaction module adopts hypernetwork to make the scenario-specific information and scenario-aware representation fully interact. Compelling results from both offline evaluation and online A/B experiments validate the effectiveness of our framework.

\bibliographystyle{splncs04}
\bibliography{reference}




% LNCS Homepage, \url{http://www.springer.com/lncs}. Last accessed 4
% \end{thebibliography}
\end{document}